\documentclass[pdflatex,sn-basic,twocolumn]{sn-jnl}

\usepackage{graphicx}
\usepackage{multirow}
\usepackage{amsmath,amssymb,amsfonts}
\usepackage{amsthm}
\usepackage{mathrsfs}
\usepackage[title]{appendix}
\usepackage{xcolor}
\usepackage{textcomp}
\usepackage{manyfoot}
\usepackage{booktabs} 
\usepackage{algorithm}
\usepackage{algorithmicx}
\usepackage{algpseudocode}
\usepackage{listings}
\usepackage{subfigure}
\usepackage{caption}  
\usepackage{orcidlink}
\usepackage{hyperref} 

\usepackage{breakurl}

\theoremstyle{definition}
\begin{document}

\title[Article Title]{Privacy-Preserving Intrusion Detection in Software-defined VANET using Federated Learning with BERT}


\author*[1]{\fnm{Shakil Ibne} \sur{Ahsan}\orcidlink{/0009-0001-9380-0079}}\email{shakil2.ahsan@live.uwe.ac.uk}

\author[2]{\fnm{Phil} \sur{Legg}\orcidlink{/0000-0003-3460-5609}}\email{phil.Legg@uwe.ac.uk}

\author[3]{\fnm{S M Iftekharul} \sur{Alam} \email{s.m.iftekharul.alam@intel.com}}


\affil[*1,2]{\orgdiv{Department of Computer Science and Creative Technologies}, \orgname{University of the West of England}, \orgaddress{\city{Bristol}, \country{UK}}}
\affil[3]{\orgdiv{Intel Labs}, \orgname{Intel Corporation}, \orgaddress{\city{California}, \country{USA}}}


\abstract{
The absence of robust security protocols renders the VANET (Vehicle ad-hoc Networks) network open to cyber threats by compromising passengers and road safety. Intrusion Detection Systems (IDS) are widely employed to detect network security threats. With vehicles' high mobility on the road and diverse environments,  VANETs devise ever-changing network topologies, lack privacy and security, and have limited bandwidth efficiency. The absence of privacy precautions, End-to-End Encryption methods, and Local Data Processing systems in VANET also present many privacy and security difficulties. So, assessing whether a novel real-time processing IDS approach can be utilized for this emerging technology is crucial. The present study introduces a novel approach for intrusion detection by the capabilities of Federated Learning (FL) in conjunction with the BERT (Bidirectional Encoder Representations from Transformers) model for sequence classification (FL-BERT). The significance of data privacy is duly recognized. According to FL methodology, each client has its own local model and dataset. They train their models locally and then send the model's weights (or parameters) to the server. After aggregation, the server aggregates the weights from all clients to update a global model. After aggregation, the global model's weights are shared with the clients. This practice guarantees the secure storage of sensitive raw data on individual clients' devices, effectively protecting privacy. After conducting the federated learning procedure, we assessed our models' performance using a separate test dataset.
The FL-BERT technique has yielded promising results, opening avenues for further investigation in this particular area of research. Also examined were several conventional machine learning models such as Random Forest (RF), Support Vector Machine (SVM), Logistic Regression (LR), and K Nearest Neighbours (KNN) to detect threats. However, these models lack data privacy. We used the Vehicular Reference Misbehavior (VeReMi) benchmark dataset in both cases. We reached the result of our approaches by comparing existing research works and found that FL-BERT is more effective for privacy and security concerns. Overall, our results suggest that FL-BERT is a promising technique for enhancing attack detection effectiveness, whereas the FL-BERT approach performs satisfactorily in attack detection and safeguarding data.}

\maketitle

\section*{Introduction}\label{sec1}

Vehicle ad-hoc Networks (VANETs) constitute a specialized subset of Mobile Ad-hoc Networks (MANETs), characterized by their dynamic topology due to high node mobility. The integration of wireless communication and computing technology in contemporary vehicles has propelled research, standardization, and development efforts towards enhancing inter-vehicle communication. This evolution is pivotal for the future of automobiles and road traffic systems, emphasizing reliance on wireless networks and machine learning to bolster road safety and mitigate highway fatalities.

The fundamental premise of VANETs lies in enabling vehicles to engage in intelligent communication, facilitating the exchange of critical information regarding road conditions and potential diversions. IEEE 802.11p serves as a communication standard in VANETs, providing a framework for vehicles to share crucial information. The architecture of Software-Defined Networking (SDN)-based VANETs, as proposed by Ku et al. \cite{Ku2014}, recognizes and addresses challenges arising from frequent disconnections, dynamic topology, and substantial node mobility within these networks. However, the heightened dynamism of VANET mobility surpasses that of MANETs, necessitating a robust mechanism to manage security vulnerabilities that could jeopardize the system.

Integrating the Software-Defined Networking (SDN) concept into VANETs has garnered increasing attention, particularly in implementing learning-based intrusion detection mechanisms. Despite making strides, Intrusion Detection Systems (IDS) in VANETs are nascent, confronting various challenges and obstacles that warrant thorough investigation. This work identifies key elements and activities requiring attention from the research community.

The literature review illuminates notable gaps in learning-based IDS research, signalling opportunities for refining IDS model accuracy while addressing data privacy and security concerns. Despite advancements, users of IDS systems continue to grapple with detecting known unencrypted attacks, with achieving 100\% accuracy remaining an elusive goal. The challenges posed by false positives and negatives underscore the imperative for optimization in IDS to mitigate these issues effectively.

In the realm of VANETs and IDS, concerns about adversarial attacks on machine learning models are prominent, whereby the deliberate injection of erroneous or malicious data may compromise the functionality of these models. Additionally, the absence of a publicly available VANET database with recorded attacks hampers progress, and the quest for a fast, cost-effective, and efficient privacy-preserving machine learning-based IDS remains an ongoing pursuit.

The communication requirements of VANETs focused on data transmission between vehicles, raise pertinent privacy concerns as sensitive information may be vulnerable. Striking a delicate balance between the imperative for continuous data exchange and preserving user privacy becomes paramount. The perpetual transmission of position data introduces privacy challenges, with the potential to misuse information about an individual's daily travel patterns. Developing an IDS system that ensures continuous monitoring with high attack detection accuracy while preventing unauthorized access or unintended data utilization presents a formidable challenge for researchers.

Exploring VANETs and IDS reveals a nuanced landscape with avenues for refinement in intrusion detection accuracy, privacy preservation, and overall system robustness. Ongoing research endeavors are indispensable for addressing these challenges and propelling the development of secure and efficient VANETs in the dynamic domain of smart transportation systems.
\smallskip

 In consequence, the contributions of this paper are as follows: 

\begin{enumerate}
    \item This paper proposes an innovative approach called FL-BERT to address privacy and security issues in the Software-defined VANET that combines the capabilities of Federated Learning (FL) with the BERT model to enhance sequence classification tasks. 
    \item In addition, conventional machine learning models were trained, particularly Random Forest (RF), Support Vector Machine (SVM), Logistic Regression (LR), and K-Nearest Neighbours (KNN) to compare evaluation reports with FL-BERT to measure effectiveness in terms of attack detection. To demonstrate that FL-BERT can perform better despite its complex structure to maintain data privacy without compromising the model's performance. 
    \item The approaches employed in our study have been evaluated using the VeReMi dataset, characterized by its substantial size and comprehensive level of information. Utilizing this dataset enables us to enhance our comprehension of the efficacy of our security techniques in detecting diverse security risks and employing privacy. This comprehensive analysis of our methodologies allows us to identify our approach's merits while highlighting potential areas for enhancement.
    \item Our research also discovers the contextual capabilities of BERT significantly improve threat detection, outperforming conventional models in terms of adaptability and performance.

\end{enumerate}

\section*{Related Works}\label{sec3}
In the course of this research, various articles have been scrutinized to elucidate the approaches employed by earlier researchers in developing an Intrusion Detection System (IDS) for detecting cyber-attacks in Software-Defined Networking (SDN)-based Vehicular Ad Hoc Networks (VANETs). The literature review uses the PRISMA methodology, encompassing eight key steps: User needs assessment, defining Research Questions, Search String formulation, Source Selection, Relevant Paper Selection, Screening Findings, Data Extraction, and data synthesis \cite{Kitchenham2007}. To discern user needs, understanding the requirements of government and private organizations, the major stakeholders in cybersecurity is crucial. Private sector involvement includes various companies, Internet Service Providers (ISPs), software vendors, and end-users (among others).

\begin{table}[!h]
\caption{VeReMi Dataset attacks and size.}
\begin{tabular}{@{}ll@{}}           \\ \hline
Attacks                                & Size   \\ \hline
BENIGN                                 & 437429 \\
Attack type 16 (Eventual stop Attack)  & 56595  \\
Attack type 4 (Random Attack)          & 30510  \\
Attack type 2 (Constant Offset Attack) & 30473  \\
Attack type 1 (Constant Attack)        & 30473  \\
Attack type 8 (Random Offset Attack)   & 29460  \\
\hline
\end{tabular}
\end{table}

Despite advancements, IDS and Malware detection systems struggle to achieve 100\% accuracy, particularly in detecting unknown, encrypted, and inference attacks. Kaspersky reported that most Malware detection products fall short of the required 90\% security threshold. Cisco, a significant player in the IDS market, emphasized the importance of rapid detection and long-term learning capabilities. However, even efficient solutions like Cisco's Encrypted Traffic Analytics have limitations, such as small attack sizes and time delays.

Creating an adaptable and versatile IDS using Machine Learning (ML) in SDN-based VANET networks faces challenges in data classification. Robust datasets are essential for precisely evaluating intrusion detection systems, with the VeReMi dataset being a notable benchmark. However, many synthetic datasets researchers generate lack size, as indicated in the table.

Several ML-based approaches are discussed in the literature. Singh (2018) proposed a method for detecting Distributed Denial of Service (DDoS) attacks in SDN-based vehicular networks using machine learning. Aneja et al. (2018) developed an AI-based system for identifying Route Request (RREQ) flooding attacks, achieving 99\% accuracy. Yu et al. (2018) presented a scheme for detecting DDoS attacks in SDVN environments using support vector machine classification. Zeng et al. (2018) employed ML-based techniques for global and local Intrusion Detection Systems (IDS) in VANETs.

Gad et al. (2021) experimented with ML approaches for intrusion detection in VANETs using the ToN-IoT dataset. Liu et al. (2014) suggested employing data mining methods in VANETs to identify existing and undiscovered threats using Naive Bayes and Logistic Regression classifiers. Türkoğlu (2022) achieved a 99.35\% accuracy score with feature selection and hyperparameter optimization using a Decision Tree classifier.

Kim (2017) proposed a collaborative security threat detection method within a software-defined vehicular cloud architecture, utilizing multiclass SVM. Misra (2011) presented a location anonymity-based system for IDS in VANET, protecting users' identities through dynamically assigned identifiers. Tian (2010) proposed a BUSNet-based IDS to protect VANETs from malicious activities using hierarchical anomaly detection.

Mejri (2014) introduced a detection system using statistical methods and linear regression to identify greedy behavior in the MAC layer passively. Alheeti (2015) developed an ANN-based intrusion detection system that identified DoS attacks with an accuracy of 85.02\% for normal behaviors and 98.45\% for deviant behaviors. Gruebler (2015) advocated using a POS approach to reduce feature dimensions for an IDS using an artificial neural network.

Sedjelmaci (2015) presented a cluster-based IDS to defend against various attacks in VANETs, utilizing detection managers at different levels and a global decision system hosted on roadside units (RSUs). The proposed system showed detection rates ranging from 92\% to 100\% depending on the number of cars and the type of attack conducted.

The literature reveals a diverse range of ML and AI-based approaches for developing IDS in SDN-based VANETs. Challenges persist, including the need for robust datasets and addressing specific attack scenarios. The research community continues to explore innovative techniques to enhance the detection capabilities of IDS in the dynamic and vulnerable VANET environment.

\begin{table*}[tbp]
\centering
\fontsize{5}{5}\selectfont
\caption{Review of different ML-based IDS models in VANET and SDVN.}
\begin{tabular}{cccccc}
\hline
Reference &
  ML Models &
  Dataset &
  Attack &
  Featuring Method &
  Top Result\\ \hline
\begin{tabular}[c]{@{}c@{}}\cite{Singh2018}\end{tabular} &
  \begin{tabular}[c]{@{}c@{}}GB, RF, DT, \\ Linear SVM, \\ LR, NN,\\ Neural Net.\end{tabular} &
  \begin{tabular}[c]{@{}c@{}}Generated \\ (14456)\end{tabular} &
  \begin{tabular}[c]{@{}c@{}}Short-lived \\ TCP Flood, \\ UDP Flood.\end{tabular} &
  - &
  \begin{tabular}[c]{@{}c@{}}Not measurable \\ from the chart\end{tabular} \\
\begin{tabular}[c]{@{}c@{}}\cite{Aneja2018}\end{tabular} &
  ANN &
  \begin{tabular}[c]{@{}c@{}}NS2 Trace \\ (1000)\end{tabular} &
  RREQ Flood &
  GA based &
  \begin{tabular}[c]{@{}c@{}}Accuracy: 95\%, \\ Precision: 97\%, \\ F-Score: 98\%, \\ FPR: 3\%\end{tabular} \\
  \\
\begin{tabular}[c]{@{}c@{}}\cite{Yu2018}\end{tabular} &
  SVM &
  \begin{tabular}[c]{@{}c@{}}DARPA 2000, \\ CAIDA \\ DDoS 2007\end{tabular} &
  \begin{tabular}[c]{@{}c@{}}TCP ﬂood, \\ UDP ﬂood, \\ ICMP Flood,\\ Etc.\end{tabular} &
  \begin{tabular}[c]{@{}c@{}}Flow-based \\ features \\ and statistical\\ based entropy\end{tabular} &
  \begin{tabular}[c]{@{}c@{}}TCP Traffic \\ Pre-Selection: \\ DR 98.26 \%, \\ FR 0.52\% \\ After-Selection: \\ DR 98.56\% \\ FR 0.32\%\end{tabular} \\
  \\
\begin{tabular}[c]{@{}c@{}}\cite{Zeng2019}\end{tabular} &
  \begin{tabular}[c]{@{}c@{}}DeepVCM, \\ KNN, DT, \\ 1D-CNN, LSTM\end{tabular} &
  \begin{tabular}[c]{@{}c@{}}ISCX2012, \\ NS3 Trace \\ (28236)\end{tabular} &
  \begin{tabular}[c]{@{}c@{}}Brute Force,\\ SSH DDoS,\\ HttpDoS\end{tabular} &
  RAW data &
  \begin{tabular}[c]{@{}c@{}}NS3 Dataset/\\ DeepVCM/DoS \\ Precision 98.5\% \\ Recall 98.3\% \\ F1-Score 98.4\%\end{tabular} \\
  \\
\begin{tabular}[c]{@{}c@{}}\cite{Zeng2018}\end{tabular} &
  ANN &
  Generate &
  - &
  \begin{tabular}[c]{@{}c@{}}CNN \\ algorithm \\ applied to \\ Extract features.\end{tabular} &
  \begin{tabular}[c]{@{}c@{}}Accuracy: \\ SMV-CASE 98.7\%, \\ CEAP 98.9\%, \\ Senior2Local 98.37\%\end{tabular} \\
  \\
\begin{tabular}[c]{@{}c@{}}\cite{Gad2021}\end{tabular} &
  \begin{tabular}[c]{@{}c@{}}LR, NB, \\ DT, SVM, \\ kNN, RF, \\ AdaBoost, \\ XGBoost\end{tabular} &
  ToN-IoT &
  \begin{tabular}[c]{@{}c@{}}DoS\\ DDoS\\ etc.\end{tabular} &
  \begin{tabular}[c]{@{}c@{}}Features \\ Selection: \\ a. Apply Chi² \\ b. Apply \\ SMOTE \\ c. Apply above \\ both\end{tabular} &
  \begin{tabular}[c]{@{}c@{}}With All Features\\  XGBoost: \\ Accuracy 99.1\%, \\ Precision 98.4\%, \\ Recall 99.1\%, \\ F1-Score 98.7\%\end{tabular} \\
  \\
\begin{tabular}[c]{@{}c@{}}\cite{Türkoğlu2022}\end{tabular} &
  \begin{tabular}[c]{@{}c@{}}kNN, SVM\\ DT\end{tabular} &
  Generated &
  DDoS &
  \begin{tabular}[c]{@{}c@{}}MRMR \\ filter \\ method\end{tabular} &
  \begin{tabular}[c]{@{}c@{}}DT Algorithm with \\ 25 Features: \\ Accuracy 99.35\%, \\ Sensitivity 99.22\%, \\ Specificity 99.80\%, \\ F1-Score 99.21\%\end{tabular} \\
  \\
\begin{tabular}[c]{@{}c@{}}\cite{Kim2017}\end{tabular} &
  SVM &
  KDD CUP 1999 &
  \begin{tabular}[c]{@{}c@{}}DoS, Probing, \\ U2R, R2L \\ Etc.\end{tabular} &
  - &
  \begin{tabular}[c]{@{}c@{}}Under Diff. Attack \\ (Approx. from \\ Graph when $\alpha$=0): \\ Accuracy \\ SVM-VC 68.00\%, \\ SVM Nearest \\ Neighbour 57.00\%, \\ SVM Individual 39.50\%\end{tabular} \\
  \\
\begin{tabular}[c]{@{}c@{}}\cite{LIU2014}\end{tabular} &
  NB, LR &
  TCPdump &
  \begin{tabular}[c]{@{}c@{}}DoS,   R2L, \\ U2R, Probing\end{tabular} &
  - &
  \begin{tabular}[c]{@{}c@{}}Not measurable \\ from Chart. \\ LR showed the \\ best result.\end{tabular} \\
\begin{tabular}[c]{@{}c@{}}\cite{Alheeti2015}\end{tabular} &
  ANN &
  NS2 Trace ﬁle &
  DoS &
  - &
  Accuracy 85.02\% \\
  \\
\begin{tabular}[c]{@{}c@{}}\cite{Gruebler2015}\end{tabular} &
  ANN &
  \begin{tabular}[c]{@{}c@{}}NS2 Trace ﬁle and\\ Animator (32000)\end{tabular} &
  Black Hole &
  - &
  Accuracy 85.02\% \\
  \\
\begin{tabular}[c]{@{}c@{}}\cite{Alheeti2016}\end{tabular} &
  \begin{tabular}[c]{@{}c@{}}FFNN,\\ SVM\end{tabular} &
  NS2 Trace ﬁle &
  \begin{tabular}[c]{@{}c@{}}Grey Hole,\\ Rushing\end{tabular} &
  - &
  \begin{tabular}[c]{@{}c@{}}Accuracy:\\ SVM-Abnormal \\ 99.80\%\end{tabular} \\

\begin{tabular}[c]{@{}c@{}}\cite{Sharma2018}\end{tabular} &
  SVM &
  NS2 Trace ﬁle &
  \begin{tabular}[c]{@{}c@{}}Wormhole, \\ Selective Forwarding,\\ Packet Drop\end{tabular} &
  \begin{tabular}[c]{@{}c@{}}Hybrid \\ Fuzzy \\ Multi-\\ Criteria \\ Feature\\ Selection\end{tabular} &
  \begin{tabular}[c]{@{}c@{}}Performance \\ evaluated \\ by DR, FPR, DT\end{tabular} \\
  
\begin{tabular}[c]{@{}c@{}}\cite{Kumar2015}\end{tabular} &
  \begin{tabular}[c]{@{}c@{}}Learning \\ automata\end{tabular} &
  Generated &
  Flooding, Blackhole &
  - &
  - \\
\begin{tabular}[c]{@{}c@{}}\cite{Shu2021}\end{tabular} &
  \begin{tabular}[c]{@{}c@{}}MLP (ANN), \\ BiGAN\end{tabular} &
  \begin{tabular}[c]{@{}c@{}}KDD99, \\ NSL-KDD\end{tabular} &
  \begin{tabular}[c]{@{}c@{}}DoS, U2R,\\ R2L, Probing etc.\end{tabular} &
  - &
  \begin{tabular}[c]{@{}c@{}}Accuracy 98.4\% \\ Precision 95.23\% \\ Recall 96.74\% \\ F1-score 95.98\% \\ Precision and recall \\ are low due to \\ complex model.\end{tabular} \\
\begin{tabular}[c]{@{}c@{}}\cite{Bangui2022}\end{tabular} &
  \begin{tabular}[c]{@{}c@{}}Hybrid model \\ based on RF\end{tabular} &
  \begin{tabular}[c]{@{}c@{}}KDD99, \\ CICIDS2017\end{tabular} &
  \begin{tabular}[c]{@{}c@{}}DoS, DDoS, \\ HeartBleed Etc.\end{tabular} &
  - &
  Accuracy 96.93\% \\
  \\
\begin{tabular}[c]{@{}c@{}}\cite{Polat2020}\end{tabular} &
  Softmax &
  \begin{tabular}[c]{@{}c@{}}Generate\\ (17779)\end{tabular} &
  DDoS &
  - &
  Accuracy 96.9\% \\
\begin{tabular}[c]{@{}c@{}}\cite{Alsarhan2023}\end{tabular} &
  \begin{tabular}[c]{@{}c@{}}SVM, GA, \\ PSO, ACO\end{tabular} &
  NSL-KDD &
  \begin{tabular}[c]{@{}c@{}}DoS, Probing, \\ Unauthorized \\ access\end{tabular} &
  - &
  Accuracy 98\% \\
\begin{tabular}[c]{@{}c@{}}\cite{Gao2019}\end{tabular} &
  RF &
  \begin{tabular}[c]{@{}c@{}}NSL-KDD, \\ UNSW-NB15\end{tabular} &
  DDoS &
  - &
  Accuracy 99.95\% \\
\begin{tabular}[c]{@{}c@{}}\cite{Alladi2021}\end{tabular} &
  \begin{tabular}[c]{@{}c@{}}CNN-LSTM, \\ CNN- bidirectional \\ LSTM, LSTM, \\ Auto-encoders \\ Stacked LSTM, \\ GRU\end{tabular} &
  VeReMi &
  \begin{tabular}[c]{@{}c@{}}Various types \\ of Anomalies\end{tabular} &
  - &
  Accuracy 98\% \\
\\  \hline
\end{tabular}
\end{table*} 

\begin{figure}[htbp]
    \centering
    \includegraphics[width=0.47\textwidth, height=5cm]{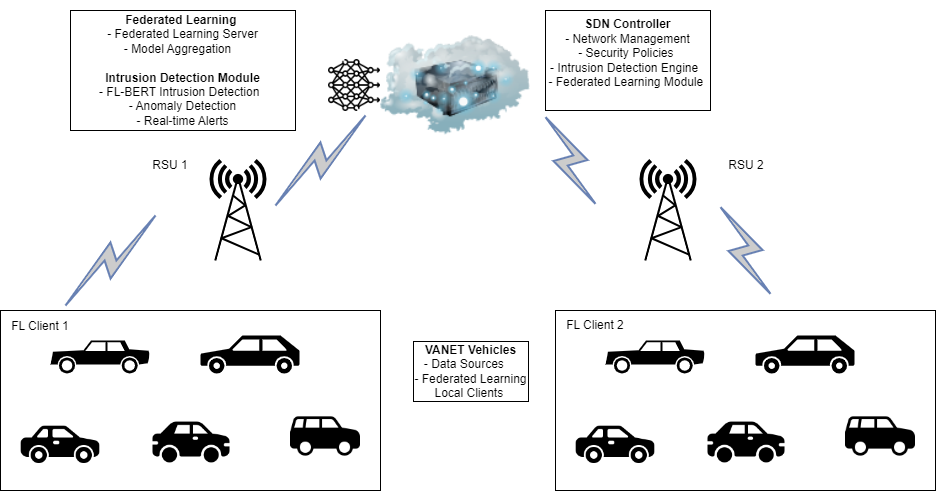}
    \caption{Overall FL-BERT model diagram.}
    \label{fig:mylabel}
\end{figure}

\section*{Methodology}\label{sec4}
In this section, we describe the selected datasets and attacks, as well as the implementation of our proposed FL-BERT model alongside conventional models, including Random Forest (RF), Support Vector Machine (SVM), Logistic Regression (LR), and K-Nearest Neighbours (KNN). All of these are intended to improve the detection of anomalies in Software-defined VANET architecture. In addition, the FL-BERT data privacy mechanisms inherent to this architecture are discussed.

The VeReMi dataset was created to test the effectiveness of VANET misbehavior detection systems (vehicular networks). This database contains onboard unit message logs created from a simulation environment and tagged ground truth. The collection contains malicious messages that are meant to cause erroneous application behavior, which misbehavior detection techniques are designed to stop. The initial dataset includes five types of position falsification attacks. \cite{van2018}
Database source: This data was generated by \cite{van2018}, and It has been collected from \cite{Sharma2021}.

To achieve the best possible hardware support for running the code, Microsoft Azure ML Studio VM instance with Python 3.8, PyTorch 1.13.1, TensorFlow 2.13.0, Protobuf 4.24.3, and Transformers 4.33.2.

\paragraph*{Random Forest Classifier}
The randomization process employed in the Random Forest Classifier is pivotal for reducing correlation, preserving robustness, and enhancing overall accuracy. In the investigated forests, each tree's growth is facilitated by applying inputs or combinations randomly selected at each node. Notably, the accuracy these forests achieve is favourable alongside Adaboost and, in some instances, even surpasses it. The Random Forest Classifier exhibits several commendable attributes:

A. Its accuracy is on par with Adaboost, occasionally surpassing it significantly.

B. Demonstrating relative resistance to noise and outliers proves to be a robust method.

C. The algorithm operates at a swifter pace compared to bagging or boosting techniques.

D. Characterized by its simplicity, ease of parallelization, and the provision of valuable core estimates encompassing error, strength, correlation, and variable relevance \cite{Breiman2001}.

\paragraph*{Logistic Regression}
Logistic Regression (LR) is a linear model primarily employed for classification tasks instead of predictive modelling. Within this framework, the logistic function is applied to ascertain probabilities of potential outcomes in a singular trial. Specifically, the logistic curve, often characterized by a shape resembling an 'S', is commonly utilized to represent this logistic function.

\[\boldsymbol{f}\left(\boldsymbol{x}\right)\boldsymbol{=}\frac{\boldsymbol{L}}{\boldsymbol{1}\boldsymbol{+}{\boldsymbol{e}}^{\boldsymbol{-}\boldsymbol{k}\left(\boldsymbol{x}\boldsymbol{-}\boldsymbol{x}\boldsymbol{0}\right)}}\] 
 \textbf{Equation 1: Logistic Function.}

Where x0 is the x value of the sigmoid's midpoint; L is the curve's maximum value; k is the curve's logistic growth rate or steepness. \cite{Wikipedia2023}

\paragraph*{Support Vector Machines}

 Support Vector Machines (SVMs) are algorithms used in learning for classification, regression, and outlier detection tasks. They are particularly effective in situations with data and can handle cases where the number of dimensions exceeds the number of samples. SVMs use a subset of training data called support vectors, which helps improve memory efficiency. One advantage of SVMs is their versatility; they can utilize kernel functions, including ones you can define yourself. However, there are some limitations to consider. Overfitting may occur if the number of features greatly exceeds the number of samples. Careful choices must be made regarding kernel selection and regularization techniques to address this issue. Additionally, SVMs do not provide probability estimates by default; obtaining these requires performing five-fold cross-validation. In the scikit learn library, SVMs can handle both sparse vectors; however, when making predictions on data, it is important first to fit the model on similar data. The decision function for an SVM is expressed as: \cite{Gad2021}

\begin{equation}
f(x) = \sum_{i=1}^{N} \alpha_i y_i K(x_i, x) + b
\end{equation}
\textbf{Equation 2: Support Vector Machines Function.}

Where \( \alpha_i \) are the Lagrange multipliers, \( y_i \) are the labels, \( K \) is the kernel function, and \( b \) is the bias.

\paragraph{K-Nearest Neighbors}
The K-Nearest Neighbors (kNN) set of rules, a case primarily based on gaining knowledge of technique, keeps all training facts for class. Its "lazy" nature limits its use in huge-scale programs like dynamic internet mining. To decorate efficiency, consultant statistics factors can be used to symbolize the whole schooling set, basically developing an inductive studying model. While numerous algorithms like selection trees and neural networks serve this cause, kNN's simplicity and effectiveness, especially obvious in textual content categorization obligations like the Reuters corpus, encourage efforts to optimize its performance without compromising accuracy. During model production, each information factor possesses a maximal local neighbourhood with statistics points of the same elegance label. The largest of these neighbourhoods, termed the "largest global neighbourhood," acts as a representative. This method repeats till all statistics points are represented. Unlike traditional kNN, this approach would not require a predefined ( k ); it is routinely decided through model production. Using representatives no longer simply reduces records but additionally boosts performance, addressing kNN's inherent barriers.

This study will address multiclass problems using the One-vs-All or OvA multiclass methodology. The steps for data preprocessing before feeding to conventional models and how to handle the FL-BERT model are discussed below.

\subsection*{Data Preprocess}

\paragraph*{A. Dataset: VeReMi}
A benchmark dataset for misbehavior detection systems for VANETs is the VeReMi dataset. It is a synthetic dataset made with LuST (Version 2) and VEINS (with modifications based on Version 4.6). In this dataset, onboard unit message logs from a simulation environment are included, along with tagged ground truth. It has two primary uses: it gives a starting point for determining how well misbehavior detection systems work city-wide. This database has very few attack classes and sizes, as shown in Table. \cite{van2018} \cite{Sharma2021}.
Table shows that we create a balanced subset of preferred samples across all attack types for model training and evaluation.

\begin{table}[!h]
\caption{VeReMi Dataset. (Resampled)}
\begin{tabular}{@{}ll@{}}           \\ \hline
Attacks                                & Size   \\ \hline
BENIGN                                 & 50000 \\
Attack type 16 (Eventual stop Attack)  & 28832  \\
Attack type 4 (Random Attack)          & 30510  \\
Attack type 2 (Constant Offset Attack) & 30473  \\
Attack type 1 (Constant Attack)        & 30473  \\
Attack type 8 (Random Offset Attack)   & 29460  \\
\hline
\end{tabular}
\end{table}

\paragraph*{B. Data Preprocessing}
We created a suitable textual feature by linking several columns from the dataset. Additionally, we encode categorical labels utilising the LabelEncoder class. The data is then split into training and testing sets, forming the basis for subsequent model development.

\paragraph*{C. FL-BERT Model}

The model being discussed in this paper is a practical application of federated learning, specifically utilising the BERT (Bidirectional Encoder Representations from Transformers) architecture. This model has been specifically tailored for use with the VeReMi dataset. Federated learning is an approach in machine learning where the training process is distributed over several devices or servers, referred to as clients, without the need to share the raw data. This approach guarantees the confidentiality of records, as the most efficient iterations of changes are transmitted to a central server, consolidating these updates to enhance a global model.

Data preparation plays a key role in the pipeline of every machine learning system. In the present model, a combination of various columns extracted from the dataset is performed, including 'sendtime', 'sender', 'messageID', 'pos', 'spd', 'sendtime', and 'sender', resulting in the creation of a unified column referred to as 'textual content'. This compiled text serves as the input for the BERT model. The column labeled 'AttackerType', potentially containing explicit data indicating unique categories of attackers, is subsequently encoded using label encoding to convert it into a numerical format. The column labelled "labels" in this encoding acts as the dependent variable for the model. The data is subsequently divided into educational and testing components, with 80\% of the data allocated for educational purposes and the remaining 20\% set aside for testing.

Tokenization is a subsequent procedure that entails transforming the factual content of the text into a structured format suitable for input into the BERT model. The utilization of the BERT tokenizer, namely the "bert-base-uncased" paradigm, serves the intended objective. To facilitate the tokenization process and prepare the records for training, a class called `CustomDataset` has been implemented. The brilliance of this approach lies in its utilization of texts, related labels, a tokenizer, and a maximum sequence time as input parameters. 

In the federated learning framework context, the model encompasses two primary components, namely the "Client" and the "Server". Every buyer corresponds to a device or server with a fraction of the training data. The user initiates a BERT model for classifying collections, transfers it to the designated computing device (GPU or CPU), and specifies an optimizer for the training process. The global model (the model aggregating updates from all local models) is continuously reset and transmitted to all clients. This means that the global model does not excessively rely on data from any specific user, which could potentially disclose patterns specific to that user. By continuously resetting the global model, it is harder for an attacker to infer specific patterns from the actions of the model, as a result enhancing privacy.

On the other hand, the server maintains a global model that consolidates the updates received from all clients. The server magnificence employs the `mixture` strategy, which involves receiving models from all clients, calculating their weights' average, and updating the global model with this computed average. This aggregation process ensures that the global model benefits from analysing individual customer models.

The federated training approach entails the process of training models on individual clients, aggregating their updates on a central server, and subsequently distributing the new global model back to the clients. The aforementioned procedure is iterated for a specific number of iterations.

The subsequent section presents an assessment function utilised to evaluate the comprehensive performance of the BERT models. This attribute sets the model to evaluation mode, which generates predictions on the test dataset and compares them with the actual labels. The performance is subsequently evaluated using a classification report, which provides metrics such as precision, recall, and F1-score for each class inside the 'AttackerType' column.

Ultimately, the proposed version presents a thorough implementation of federated learning using the VeReMi dataset's BERT architecture. This version provides a robust solution for training on distributed data while ensuring data privacy, by utilising the capabilities of BERT and the privacy-preserving characteristics of federated learning.

\paragraph*{D. Conventional Model}

In the course of our study, we performed an assessment utilizing machine learning classifiers such as Random Forest (RF), Support Vector Machine (SVM), Logistic Regression (LR), and K Nearest Neighbours (KNN). The primary objective of our study was to detect abnormalities in the 'AttackerType' within a Software Defined VANET architecture. The dataset was partitioned into training and test subsets, with the 'text' column serving as the feature.

The Term Frequency Inverse Document Frequency (TF IDF) vectorization technique was employed to preprocess the data for machine learning. The proposed methodology quantifies the significance of individual terms within the dataset by considering their frequency across all documents, resulting in a normalised representation. We have restricted the features to the highest-ranking one thousand to enhance efficiency and prioritise specific terminologies. 

We trained each classifier using the modified training dataset following the vectorization process. Subsequently, an assessment was conducted to evaluate their performance on the test dataset. By employing this methodology, we evaluated each classifier's efficacy and constraints in identifying anomalies. The study of the findings includes precision, recall, and F1 score measures, which identify the most effective models for our specific application case.

\section*{Results and Evaluation}\label{sec5}

In this study, we performed a comparative analysis of machine learning models and the innovative Federated Learning BERT (FL BERT) approach to identify abnormalities in Software-Defined Vehicular Ad Hoc Network (VANET) infrastructures. We evaluated many models, including Random Forest (RF), Support Vector Machine (SVM), Logistic Regression (LR), and K Nearest Neighbours (KNN).

The FL BERT model exhibited an 84\% accuracy rate, substantiating its efficacy in processing context-rich data. The system continuously demonstrated high levels of precision, recall, and F1 score metrics across various attacker types, suggesting its capacity to detect a wide range of anomalies accurately.

\begin{figure*}[htbp]
    \centering
    \begin{minipage}{0.5\textwidth}
        \centering
        \captionof{table}{Accuracy Comparison} 
        \begin{tabular}{lc}
            \toprule
            Models & Accuracy \\
            \midrule
            Random Forest & 49 \\
            SVM & 59 \\
            Logistic Regression  & 59 \\
            KNN & 38 \\
            FL-BERT & 84 \\
            \bottomrule
        \end{tabular}
    \end{minipage}%
    \begin{minipage}{0.5\textwidth}
        \centering
        \includegraphics[width=0.9\linewidth]{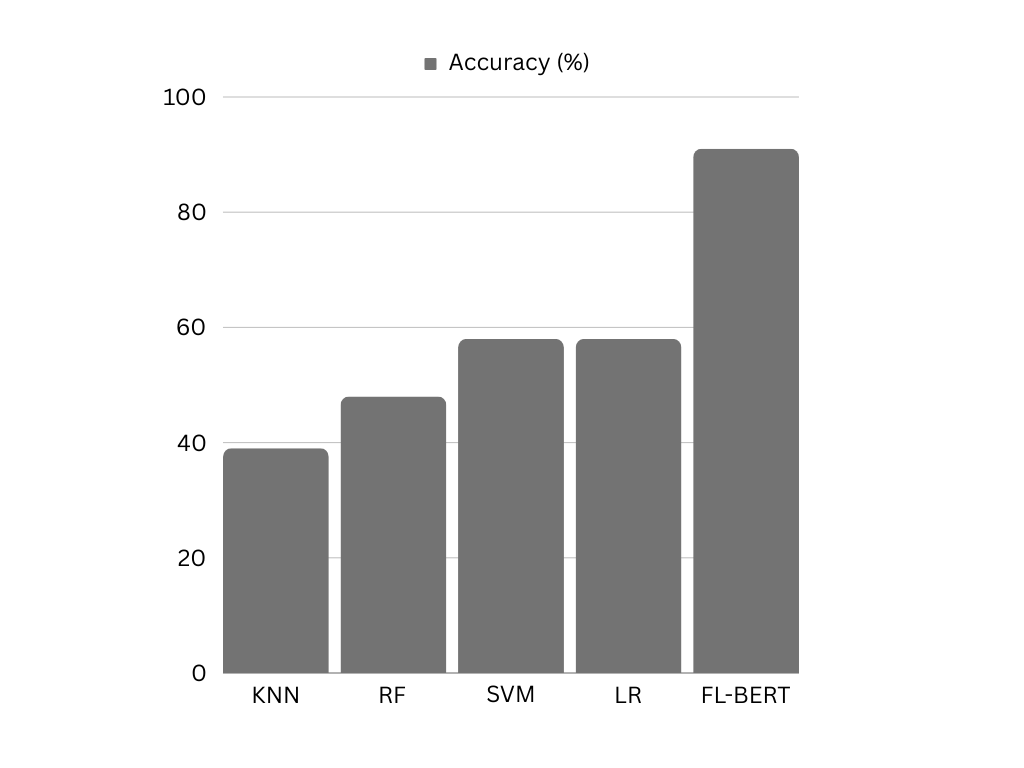}
        \caption{Accuracy comparison chart}
        \label{fig: Accuracy Chart}
    \end{minipage}
\end{figure*}

On the other hand, the Random Forest algorithm demonstrated an accuracy rate of 49\%, indicating competence in certain categories while exhibiting deficiencies in others. Support Vector Machine (SVM) and Logistic Regression models demonstrated a comparable accuracy rate of 59\%. Although these models showed proficiency in identifying certain types of attackers, their performance was comparatively poorer in other areas. This discrepancy suggests the possibility of overfitting or insufficiency in effectively addressing the intricate nature of the dataset.

\begin{table*}[htbp]
    \small
    \centering
    \caption{Performance Metrics for Different Models}
    \begin{tabular}{lcccccc}
        \toprule
        Model & Type & Precision & Recall & F1-Score & Support \\
        \midrule
        \multirow{6}{*}{FL-BERT} & 0 & 0.94 & 0.95 & 0.95 & 9899 \\
                                 & 1 & 1.00 & 1.00 & 1.00 & 6158 \\
                                 & 2 & 0.65 & 0.88 & 0.75 & 6096 \\
                                 & 4 & 0.98 & 0.74 & 0.84 & 6181 \\
                                 & 8 & 0.69 & 0.45 & 0.55 & 5899 \\
                                 & 16 & 0.79 & 0.95 & 0.86 & 5717 \\
        \midrule
        \multirow{6}{*}{Random Forest} & 0 & 0.73 & 0.79 & 0.76 & 9899 \\
                                       & 1 & 1.00 & 1.00 & 1.00 & 6158 \\
                                       & 2 & 0.35 & 0.35 & 0.35 & 6096 \\
                                       & 4 & 0.15 & 0.14 & 0.14 & 6181 \\
                                       & 8 & 0.12 & 0.11 & 0.12 & 5899 \\
                                       & 16 & 0.35 & 0.33 & 0.34 & 5717 \\
        \midrule
        \multirow{6}{*}{SVM} & 0 & 0.76 & 0.83 & 0.79 & 9899 \\
                              & 1 & 1.00 & 1.00 & 1.00 & 6158 \\
                              & 2 & 0.53 & 0.42 & 0.47 & 6037 \\
                              & 4 & 0.32 & 0.34 & 0.33 & 6190 \\
                              & 8 & 0.31 & 0.33 & 0.32 & 5892 \\
                              & 16 & 0.48 & 0.45 & 0.46 & 5784 \\
        \midrule
        \multirow{6}{*}{Logistic Regression} & 0 & 0.73 & 0.82 & 0.77 & 9899 \\
                                             & 1 & 1.00 & 1.00 & 1.00 & 6158 \\
                                             & 2 & 0.50 & 0.46 & 0.48 & 6037 \\
                                             & 4 & 0.33 & 0.37 & 0.35 & 6190 \\
                                             & 8 & 0.33 & 0.30 & 0.31 & 5892 \\
                                             & 16 & 0.48 & 0.39 & 0.43 & 5784 \\
        \midrule
        \multirow{6}{*}{KNN} & 0 & 0.65 & 0.77 & 0.70 & 9899 \\
                             & 1 & 0.52 & 0.58 & 0.55 & 6148 \\
                             & 2 & 0.21 & 0.21 & 0.21 & 6037 \\
                             & 4 & 0.13 & 0.13 & 0.13 & 6190 \\
                             & 8 & 0.13 & 0.11 & 0.12 & 5892 \\
                             & 16 & 0.35 & 0.25 & 0.30 & 5784 \\
        \bottomrule
    \end{tabular}
\end{table*}

The KNN model had a lower accuracy rate of 38\%, indicating that it may not be well-suited for the given data distribution or the intrinsic difficulties of the VANET system.
In summary, while standard machine learning models have demonstrated varying levels of efficacy, FL BERT has emerged as a viable choice for detecting abnormalities in Software-Defined VANETs. In handling various types of assailants, this model exhibits superior accuracy and precision compared to its counterparts. This study highlights the potential of utilizing advanced neural network models to address complex cybersecurity concerns.

In intrusion detection, the FL-BERT model exhibits a high degree of precision and recall for Attack type 1 (Constant Attack) and Attack type 2 (Constant Offset Attack), with F1-scores of 0.95 and 1.00, respectively. This suggests that the model is highly effective at identifying and correctly classifying these types of attacks.

For Attack type 4 (Random Attack), the model shows a relatively lower precision of 0.65 but a high recall of 0.88, resulting in an F1-score of 0.75. This indicates that while the model correctly identifies a high proportion of this attack type (high recall), it also incorrectly classifies other types as this attack (lower precision).

The model's performance on Attack type 8 (Random Offset Attack) and Attack type 16 (Eventual stop Attack) is mixed, with F1-scores of 0.55 and 0.86 respectively. The lower F1-score for Attack type 8 suggests that this type of attack is more challenging for the model to identify and classify correctly.

In conclusion, with its federated learning approach, the FL-BERT model demonstrates a promising capability in intrusion detection tasks, particularly in classifying different types of attacks. However, the varying performance across attack types suggests potential areas for further optimization and improvement. 

FL-BERT's advantages and disadvantages are similarly highlighted by evaluating using standard methods like Random Forest and SVM. Compared to FL-BERT, the much-reduced precision and do-not-forget values in these typical designs highlight how complicated the information is. Without the benefit of deep architectures, such models might not be able to grasp the subtle styles in VANET records, leading to more frequent misclassifications.

The imbalance of the records is a further important consideration. The training data structure is crucial in determining how a model makes predictions. The model's knowledge may be skewed if certain assault types predominate in the dataset, making it either too aggressive or cautious. For instance, an excess of benign statistics may make the version prone to making warning errors, leading to the classification of capability risks as unimportant. In conclusion, although FL-BERT claims outstanding accuracy, understanding these nuances and continuously improving the version based only on these insights are crucial to minimizing misclassifications.

\section*{FL-BERT Privacy Concern}
Implementing FL-BERT in a Software-defined Vehicular Ad-hoc Network (VANET) demonstrates the integration of advanced deep learning techniques with the intricate realm of vehicular communications. Vehicular Ad Hoc Networks (VANETs) are specifically engineered to introduce a novel era of intelligent transportation systems, wherein vehicles establish communication to ensure safety and optimize traffic flow. Nevertheless, the nature of statistics transmitted across these networks is sensitive. The continual sharing of information regarding every vehicle's role, speed, route, and other factors raises privacy concerns, highlighting its importance.

Introducing FL-BERT, a customized iteration of the esteemed BERT model designed for federated learning. Federated Learning (FL) is a notable technology that enables models to acquire knowledge from several devices or servers without the requirement of centralizing data. The primary advantage of employing this technique lies in its ability to provide strict localization of raw recordings, which can be highly illuminating within the context of Vehicular Ad-Hoc Networks (VANETs). Instead of transmitting raw data, only updated model versions, such as gradients or weights, are returned to a central server. This implies that intricate details regarding specific vehicles, their routes, or communication patterns are never disclosed to any significant entity or possible individuals attempting to intercept such information.

Preserving anonymity is more appropriate when a central server collects updates from multiple clients to enhance the global model. The process of aggregating automatically provides a level of anonymization. By amalgamating updates from multiple sources, the task of reverse engineering or deducing specific details about an individual information source or user will become exceedingly challenging. 

The periodic re-initialization of the version represents a widely implemented yet significant measure to enhance privacy. The gadget employs regularly resetting and disseminating the global version to all clients, preventing overfitting and excessive reliance on data from any specific user. The aforementioned behavior is a deterrent for potential attackers who may attempt to infer specific patterns from the actions of the version in question.

The significance of BERT, characterized by its transformer architecture, is equally crucial. Transformers exhibit a high level of proficiency in processing sequential data and have the capability to capture intricate patterns. In the context of Vehicular Ad-Hoc Networks (VANETs), this implies that the proposed model can effectively process vehicular data streams without the need for explicit feature engineering. This enhances the model's accuracy and mitigates the potential risk of information leakage occasionally occurring with manually designed routines.

Nevertheless, it is important to note that no technology is completely impervious to vulnerabilities. Despite implementing federated studying, the persistence of vulnerabilities, like version inversion attacks, remains a concern. These attacks include malicious entities attempting to reconstruct raw information from version updates. To enhance the security measures of the gadget, incorporating additional techniques, such as differential privacy, would prove advantageous.

Using FL-BERT in VANETs is a significant step towards a future in which vehicular networks are more intelligent and prioritize each participant's privacy. The results, which highlight the superior performance of FL-BERT compared to conventional models such as Random Forest and SVM, also underscore the potential of federated learning in these significant applications.

\section*{Conclusion }
The need for robust security and privacy solutions is increasingly emphasized in the dynamic landscape of Software-defined Vehicular Ad Hoc Networks (VANETs). This study has made significant contributions in shedding light on a promising path forward, offering contemporary solutions that effectively negotiate the complex interplay between advanced machine learning and data privacy. The emergence of FL-BERT, a framework that combines Federated Learning with the BERT model, exemplifies the revolutionary potential of integrating decentralized learning with advanced text categorization algorithms. This integration not only effectively handles the urgent challenges of data privacy but also ensures that the effectiveness of risk detection is not affected.

The superiority of FL-BERT is further highlighted through our comparative evaluation with traditional machine learning models such as RF, SVM, LR, and KNN. FL-BERT consistently demonstrates impressive performance metrics, outperforming its more conventional counterparts, despite its intricate structure that prioritizes data privacy. The aforementioned findings demonstrate a significant shift in paradigms, implying that the future of secure VANETs may rely on using federated learning frameworks that exploit advanced device learning architectures.

Including the VeReMi dataset in our research has enhanced the severity of our empirical examination. The comprehensive scope of our study facilitated a thorough examination of our proposed methodologies, yielding valuable, enduring, illuminating insights. The dataset functioned as a testing ground, evaluating the effectiveness of our approaches against a wide range of security risks and, in doing so, identifying areas of strength and potential improvement.

Moreover, the observable scope facilitated by BERT's contextual abilities reinforces the increasing agreement within the academic community about the revolutionary role of transformer-based models in threat detection. The adaptability and overall performance of the subject, particularly when compared to conventional models, provide a promising direction for future study efforts.

This study represents a significant advancement in pursuing secure and privacy-preserving Vehicular Ad Hoc Networks (VANETs). By advocating for implementing federated learning and leveraging advanced machine learning models, we present a comprehensive framework that enhances security and demonstrates a strong dedication to safeguarding data privacy. The insights and approaches presented in this study are expected to be crucial in shaping future advancements in VANETs. This study basically recommends implementing this solution to an SDN controller in VANET. \cite{Banitalebi2021}

\section*{Declarations}
\subsubsection*{Competing interests}
The authors declare that they have no competing interests.
\subsubsection*{Authors' contributions}
Work equally conducted by Shakil Ibne Ahsan and Co-authors. 

\subsubsection*{Availability of data and materials}
Not applicable. 
\subsubsection*{Funding}
Not applicable. 
\subsubsection*{Acknowledgements}
The genuine gratitude goes to the FET-Computer Science and Creative Technologies department, University of the West of England - UWE Bristol, for providing essential, necessary support to enrich this study. The authors would like to thank the reviewers for their valuable time. \medskip

\begin{appendices}

\end{appendices}

\bibliography{sn-bibliography}

\end{document}